\documentclass[10pt,letterpaper]{article}
\usepackage[top=0.85in,left=2.75in,footskip=0.75in]{geometry}

\usepackage{amsmath,amssymb}

\usepackage{changepage}

\usepackage[utf8x]{inputenc}

\usepackage{textcomp,marvosym}

\usepackage{cite}

\usepackage{hyperref}
\usepackage[right]{lineno}

\usepackage{microtype}
\DisableLigatures[f]{encoding = *, family = * }

\usepackage[table]{xcolor}

\usepackage{array}

\usepackage{subfigure}

\newcolumntype{+}{!{\vrule width 2pt}}

\newlength\savedwidth

\newcommand\thickhline{\noalign{\global\savedwidth\arrayrulewidth\global\arrayrulewidth 2pt}%
\hline
\noalign{\global\arrayrulewidth\savedwidth}}

\raggedright
\usepackage[aboveskip=1pt,labelfont=bf,labelsep=period,justification=raggedright,singlelinecheck=off]{caption}

\makeatletter
\renewcommand{\@biblabel}[1]{\quad#1.}
\makeatother

\date{}

\usepackage{lastpage,fancyhdr,graphicx}
\usepackage{epstopdf}
\fancyheadoffset[L]{2.25in}
\fancyfootoffset[L]{2.25in}

\begin{document}
\vspace*{0.2in}

\begin{flushleft}
{\Large
\textbf\newline{Vector-borne disease risk indexes in spatially structured populations}
}
\newline
\\
Jorge Vel\'azquez-Castro \textsuperscript{1*}, 
Andr\'es Anzo-Hern\'andez \textsuperscript{2},
Beatriz Bonilla-Capilla      \textsuperscript{2}, 
Mois\'es Soto-Bajo              \textsuperscript{2},
Andr\'es Fraguela-Collar     \textsuperscript{1}. 
\\
\bigskip
\textbf{1} Facultad de Ciencias F\'isico-Matem\'aticas, Benem\'erita Universidad Aut\'onoma de Puebla, Avenida San Claudio y 18 Sur, Colonia San Manuel, 72570. Puebla, Puebla, M\'exico.\\ 
\textbf{2} C\'atedras CONACYT - Benem\'erita Universidad Aut\'onoma de Puebla - Facultad de Ciencias F\'isico-Matem\'aticas, Benem\'erita Universidad Aut\'onoma de Puebla, Avenida San Claudio y 18 Sur, Colonia San Manuel, 72570. Puebla, Puebla, M\'exico.

%
%





* jorgevc@fcfm.buap.mx

\end{flushleft}
\section*{Abstract}
There are economic and physical limitations when applying prevention and control strategies for urban vector borne diseases. Consequently, there are increasing concerns and interest in designing efficient strategies and regulations that health agencies can follow in order to reduce the imminent impact of viruses like Dengue, Zika and Chikungunya. That includes fumigation, abatization, reducing the hatcheries, picking up trash, information campaigns. A basic question that arise when designing control strategies is about which and where these ones should focus. In other words, one would like to know whether preventing the contagion or decrease vector population, and in which area of the city, is more efficient. 

In this work, we propose risk indexes based on the idea of secondary cases from patch to patch.  Thus, they take into account human mobility and indicate which patch has more chance to be a corridor for the spread of the disease and which is more vulnerable, \textit{i.e.} more likely to have cases?. They can also indicate the neighborhood where hatchery control will reduce more the number of potential cases.

In order to illustrate the usefulness of these indexes, we run a set of numerical simulations in a mathematical model that takes into account the urban mobility and the differences in population density among the areas of a city. If we label by $i$ a particular neighborhood, the transmission risk index ($TR_{i}$) measures the potential secondary cases caused by a host in that neighborhood. The vector transmission risk index ($VTR_{i}$) measures the potential secondary cases caused by a vector. Finally, the vulnerability risk index ($VR_{i}$) measures the potential secondary cases in the neighborhood. Transmission indexes can be used to give geographical priority to some neighborhoods when applying prevention and control measures. On the other hand, the vulnerability index can be useful to implement monitoring campaigns or public health investment. 

\section*{Author summary}
A disease is called vector-borne when it is not transmitted directly among humans, but in a human-vector-human way. Examples of major importance, due to its epidemiological magnitude, are Dengue, Chikungunya and Zika tropical diseases, for which the main vector is the \textit{Aedes aegypti} mosquito. Usually, some indexes are used to measure the potential damage that such a disease could cause, which concern infectivity and entomological issues, human mobility and geographical density, etc. 

We consider a mathematical epidemiological model which takes specifically into account these spatial factors (mobility, density), and propose a set of new risk indexes which give information about how the epidemic spread could occur, and where the outbreaks could take place. These indexes come from a preventive perspective, and they also pay attention to the epidemic dynamics. Consequently, they potentially allow to proceed in the correct places, and before troubles have arisen. We think these new tools could eventually help health agencies in the design of efficient and effective strategies on prevention and control of vector-borne diseases epidemics. 


\section*{Introduction}

Dengue, Zika and Chikungunya are deseases of major concern in many countries \cite{Manore2014,Bichara-Holechek-JVelazquez-2016}. In particular, Dengue is of major importance due to its epidemiological magnitude \cite{Jansen2010,Kraemer2015}. On the other hand, in many countries Zika and Chikungunya can be considered as new diseases for which a great portion of the population is susceptible. This feature make them a major epidemiological threat. 

These three deseases are vector borne, transmitted by the bite of \textit{Aedes aegypti} mosquito \cite{Ferreira2006,Manore2014}. Thus the greatest efforts in prevention and control are in reducing the mosquito population. This can be done by spying or by reducing the hatcheries. The \textit{A. aegypti} is a domestic species in the sense that it oviposit its eggs in human made containers with clean water, so it is principally found around houses \cite{Jansen2010a,Kraemer2015}. The containers used by the mosquitoes for reproducing can be trash in yards or streets filled with water from rain, drums that some people use to accumulate water in case it goes scarce, or even pools and swimming pools \cite{Reisen2010}. In addition to fumigation, other strategies has been implemented to control vector reproduction capacity, like picking up unprocessed trash, abatization... The implementation of information campaigns promoting to cap water recipients, the use of repellents and mosquito nets, or even visit the physician at first symptoms have been also important. 

Implementing this measures cost money, so in big to medium cities is not normally possible to apply them in all the urban area, and a lot less in a whole country. Thus, local governments are forced to take a decision about which measures to implement with the resources at hand and where they are going to be implemented \cite{Gratz2004}. Thus, in the process to better manage the resources they always face the following questions: Where should control measures be implemented? Which kind of measures should be implemented in order to get the best impact in controlling the dispersion of the disease? 

Nowadays, the traditional control and preventive measures are based on population indexes \cite{Ames2011,Ferreira2006,Sanchez2006}. That means that if the larvae density is higher than a certain empirically pre-established threshold, alert is raised, and in case there are enough resources, hatchery control is carried out. These traditional indexes used by entomologists have the following disadvantages. First, they are reactive and not preventive, because of the measures are taken once a high density of mosquitoes population is detected and nothing is done to prevent its growth before that. Second, even if vector density is one of the main factors promoting the disease transmission, this kind of indexes overlook other important factors like human density and mobility \cite{Bansal2007a,Stoddard2009}. 

Human mobility and density play an important role not only in outbreaks but also in the maintenance of endemic estates 
\cite{Almeida2010,Aparicio2007,Castillo-Bichara2016,Cosner2009,Stoddard2009}. Thus, either an strategy to reduce the impact of an outbreak or an strategy to eliminate an endemic situation should take into account these human factors \cite{Bichara2015,Diekmann1990}. This is also true because of these kind of indexes do not take into account the dynamics of the disease that is determined by the transmission mechanisms 
\cite{Apolloni2014,Bichara-CastilloChavez2016,Colizza2008,Revilla2008}. 

In this work, we derive risk indexes based in the idea of secondary cases. Then, we obtain explicit expressions for them in the frame of a dynamical mathematical model for the spread of \textit{Aedes aegypti} borne deseases. We propose a model within a meta-population framework with a Lagrangian approach in order to take into account human intra-urban mobility. The proposed indexes arise in a natural way from the dynamical model using the theory of dynamical systems and networks \cite{Keeling2011,Pastor-Satorras2015,Aparicio2007,Eubank2010,Molina2012}. 

We focus our attention in two types of indexes. One of them, is a measure of the potential transmission and spread of the disease caused by human behavior. The other kind of index is a measure of the transmission caused by the particular spatial distribution of mosquitoes. Thus, each kind of index is useful to determine the most effective type of strategy, either in avoiding human contagion and prevention, like campaigns to go to hospitals at first symptoms, or avoiding mosquito effects, like abatization, fumigation and hatchery elimination. 
Even if this indexes can be used to guide control measures they are build to guide prevention measures, thus they are independent of an outbreak state. This is a main difference between the proposed risk indexes and optimal control strategies where the state of the epidemics needs to be known thus being reactive and not preventive.

While the proposed indexes consider population densities, they also take into account human mobility, so they can be used to propose a global control strategy. That means that the risk indexes take into account that a neighborhood with important human inward flux and outward flux has more chance to be a corridor for the spread of the disease. Thus, the kind of questions that can be answered using these indexes are: Which area is more likely to act as a corridor and which is more vulnerable, \textit{i.e.} more likely to have cases? In which neighborhood the hatchery control will reduce more the number of potential cases? In which neighborhood will the resource allocation be more effective in surveillance of cases and immediate attention? 

\section*{Materials and methods}

\subsection*{Risk indexes}

We look for to define summary measures that capture important information of how a virus transmitted by a vector (like the \textit{A. aegypty} mosquito) is spread in a region. Let $\Omega$ be the spatial region where the vector-borne disease transmission will be analyzed. Such region could be a city or a village which can be divided into $N$ disjoint subregions $\Omega_i$, such as neighborhoods, which we call patches. Thus, $\Omega=\cup_{i=1}^{N}\Omega_i$. Hosts and vectors coexist, but each patch has its particular social and ecological features, \textit{i.e.} human population, transportation and mobility habits of the inhabitants, and mosquitoes density can differ from one neighborhood to other. 

Because of the \textit{A. aegypty} has a very limited range of flight, the virus is mainly spread among patches just by human mobility. In general, the spread process caused by human mobility is as follows. An individual that normally is in patch $i$ (a resident) travels to patch $k$, where they becomes infected by a mosquito. Then they travels again to an other patch $j$, where a mosquito there is infected by them. The result is that the virus was spread from patch $k$ to patch $j$ by a resident from $i$. We can illustrate the generality of this with two particular cases. When $i=j$, it corresponds to an individual introducing the disease in their own patch because of they got infected in a different patch. On the other hand, $i=k$ represents a case when an already infected individual travel to a different patch and spread the virus in that particular patch. 

The secondary cases caused by a single infected individual at a completely susceptible population, normally called the basic reproduction number $R_{0}$, is one of the most important summary measures that describe the severity of an epidemic outbreak. In what follows we extend this idea to include geographical information of where the secondary cases where produced and which is the origin of the infected hosts. 

First we denote $R_{kj}^{(v)}$ as the secondary human infections of residents from $j$ produced by an infected vector in $k$ in the disease-free state. In a similar way, we define $R_{ik}^{(h)}$ as the number of vector secondary cases in $k$ caused by a single infected resident in $i$. These quantities can be calculated from a mathematical model or even from a particular outbreak. 

A natural way of deriving useful risk measures from these quantities is to obtain the secondary infections $R_{i}^{(h)}$ generated by a single individual of patch $i$ in the total susceptible population system:
\begin{equation}
R_{i}^{(h)} = \sum_{j=1}^{N} \sum_{k=1}^{N} R_{ik}^{(h)} R_{kj}^{(v)}\,.
\end{equation}
Here $R_{ik}^{(h)}R_{kj}^{(v)}$ is the number of human-human secondary infections that a resident of $i$ generated in residents of $j$ that were produced in patch $k$. Summation in $k$ is to take into account all the possible places where the contagion could take place. Summation in $j$ is to account for all the secondary human infections that a single resident of $i$ produces in all the system. Thus $R_{i}^{(h)}$ is a measure of the contagion capacity of residents from the patch $i$. 

On the other hand, the classic definition of risk is the probability of occurrence of an unwanted event multiplied by the consequence of the event. Following this definition, the transmission risk index $TR_{i}$ is the probability of a person of patch $i$ gets infected multiplied by the secondary cases it generates. If we denote by $N_{hi}$ the human population of patch $i$ and $N_{h}$ the system-wide population, then the Transmission risk index is given by
\begin{equation}
TR_{i} = R_{i}^{(h)}\,\frac{N_{hi}}{N_{h}}\,,
\end{equation}
where $N_{h}=\sum_{i=1}^{N}N_{hi}$ and we have assumed that the probability of finding the first infected person in a given patch, is proportional to the patch population. Thus, $TR_{i}$ indicates the risk of the neighborhood $i$ to become the main disperser of the disease at the beginning of an epidemic. 

On the other hand, the number of secondary cases of $j$ residents caused by a single infected resident of $i$ is given by
\begin{equation}
\sum_{k=1}^{N} R_{ik}^{(h)}\,R_{kj}^{(v)}\,.
\end{equation}
The summation in the previous expression accounts for the vector contagions in all patches. We can now define a Vulnerabiliy risk index $VR_{j}$ for patch $j$ as
\begin{equation}
VR_{j} = \sum_{i=1}^{N}\sum_{k=1}^{N} \frac{N_{hi}}{N_{h}}\,R_{ik}^{(h)}\,R_{kj}^{(v)}\,.
\end{equation}
The factor $\frac{N_{hi}}{N_{h}}$ in each term of the summation accounts for the probability of the initial infected host is a resident of $i$. 

Finally, we define the Vector transmission risk index $VTR_{i}$ as the secondary human infections $R_{i}^{(v)}$ caused by an infected vector in $i$ multiplied by the probability of a vector of $i$ becomes infected at the beginning of an epidemic. Denote by $N_{vi}$ the vector population of patch $i$. If we denote by the constant $w_{i}$ the time average of the host population that is at $i$ independently of where they come from, then the $VTR_{i}$ is expressed as
\begin{equation}
VTR_{i} = \frac{w_{i}}{N_{h}}\,
\left(1 - \left(1-\frac{1}{w_{i}}\right)^{N_{vi}} \right)\,
R_{i}^{(v)}\,,
\end{equation}
where $\frac{w_{i}}{N_{h}}$ is the probability of the original infected host is in $i$, $\left(1 - \left(1-\frac{1}{w_{i}}\right)^{N_{vi}} \right)$ is the probability of it gets bitten by a mosquito in $i$, and $R_{i}^{(v)}$ is calculated as
\begin{equation}
R_{i}^{(v)} = \sum_{j=1}^{N} R_{ij}^{(v)}\,.
\end{equation}
In this case $R_{i}^{(v)}$ represents the system-wide secondary human infections caused by a single infected vector in $i$. This index will be important just in the cases when the number of mosquitoes is large, so we can approximate it by
\begin{equation}
VTR_{i} \simeq \frac{w_{i}}{N_{h}}\,
\left(1 - e^{-N_{vi}/w_{i}} \right)\,R_{i}^{(v)}
\qquad {\rm for}\quad N_{vi} \gg 1\,.
\end{equation}

\subsection*{Model description}

Our model aims to capture the dynamics of a virus transmitted by vectors (like the \textit{Aedes aegypty} mosquito). Important viruses transmitted by this species are Dengue, Zika and Chikungunya. 
In order to take into account human mobility and density population within the city, we divided it into different areas or neighborhoods that we will call patches. In each patch hosts and vectors coexist, and every one has its particular social and ecological features, \textit{i.e.} human population, transportation and mobility habits of the inhabitants, and mosquitoes density can differ from one neighborhood to other. In addition, hosts travel among patches. In this sense, patches are said to be connected. 

The model is thus built in two steps. First we build a general model for a single patch, that describes the dynamics of a vector-borne disease in a single patch. After that, we extend the local model for multiple patches, representing the whole city and mobility effects. 

\subsubsection*{Single patch model}

We consider that for a given subregion $\Omega_{i}$, with $i \in \{1,2,\ldots,N\}$, hosts and vectors are homogeneously mixed. Furthermore, we assume host population can be classified in susceptible, infectious and recovered. On the other hand, vector population is classified in susceptible and infectious. Let $N_{hi}$, $S_{hi}$, $I_{hi}$ and $R_{hi}$ denote the number of total, susceptible, infectious and recovered host individuals, respectively, and analogously let $N_{vi}$, $S_{vi}$ and $I_{vi}$ denote number of total, susceptible and infectious vector individuals respectively of patch $\Omega_i$ at a certain time. Thus, we have $N_{hi}=S_{hi}+I_{hi}+R_{hi}$ and  $N_{vi}=S_{vi}+I_{vi}$ at any time. 

In table \ref{table1} we describe each parameter interpretation and its reference value. 

\begin{table}[!ht]
\begin{adjustwidth}{-2.25in}{0in} 
\caption{
{\bf Description and estimated value of the model parameters (\cite{adams2010}, \cite{esteva_modelling_2006}).}}
\begin{tabular}{|c|l|c|}
\hline
\textbf{Parameter} & \textbf{Descrition} & \textbf{Value}  \\ \thickhline
$\beta$ & Effective vector biting rate (global number of bites, per day, per mosquito) & $0.67$ \\ \hline
$\gamma$ & Recovery rate of hosts (per day, per capita) & $1/7$ \\  \hline
$\alpha_{i}$ & Number of eggs laid per day for every female mosquito in the $i$-th patch & $5$ \\ \hline
$C_{i}$ & Carrying capacity of hatcheries for adult female mosquitoes in the $i$-th patch & - \\  \hline
$\mu_{i}$ & Per-capita mortality rate of adult female mosquitoes in the $i$-th patch & $1/8$ \\ \hline
\end{tabular}
\label{table1}
\end{adjustwidth}
\end{table}
 
For the sake of simplicity, in this subsection we will omit the patch index $i$.We consider no total hosts population change ($\dot{N_{h}}=0$), which is reasonable for a short period of time in which an outbreak occurs. On the other hand, for vectors population we use a logistic population model. The carrying capacity parameter $C$ of vector population plays a major role, since it is related with the socio-environment features of the subregion that bounds vector population growth. In specific, carrying capacity is the maximal load of mosquitoes (or vectors) that environment can support in a given patch. If vital resources for the mosquitoes were unlimited, the susceptible mosquitoes $S_{v}$ would grow at a rate $\alpha$ (ignoring mortality); however, the factor $1-N_{v}/C$ takes into account the decrease of the per-capita reproductive rate of mosquitoes due to the saturation of hatcheries. 

On the other hand, the hosts population dynamics is described by a SIR type model. The coupling of both models give rise to the following system of differential equations:
\begin{eqnarray}
\dot{S}_{h} & = & 
-\beta I_{v}\,\frac{S_{h}}{N_{h}}\,, 
\label{local model eq1} \\ 
\dot{I}_{h} & = & 
\beta I_{v}\,\frac{S_{h}}{N_{h}} -\gamma\,I_{h}\,, 
\label{local model eq2} \\ 
\dot{R}_{h} & = & 
\gamma\,I_{h}\,, 
\label{local model eq3} \\ 
\dot{S}_{v} & = & 
\alpha\,N_{v}\,\left(1-\frac{N_{v}}{C}\right) 
-\beta\,S_{v}\,\frac{I_{h}}{N_{h}} -\mu\,S_{v}\,, 
\label{local model eq4} \\ 
\dot{I}_{v} & = & 
\beta\,S_{v}\,\frac{I_{h}}{N_{h}} -\mu\,I_{v}\,. 
\label{local model eq5} 
\end{eqnarray}

As usual, a susceptible vector becomes infectious when it bites an infectious host. Then, in equations \eqref{local model eq1} and \eqref{local model eq2}, $\beta\,I_{v}$ is the total number of bites of infectious vectors to host per unit of time. From the total number of bites, only a portion $S_{h}/N_{h}$ affects susceptible hosts. In equations \eqref{local model eq4} and \eqref{local model eq5}, $\beta\,S_{v}$ is the total number of bites performed by susceptible vectors, from which only a fraction $I_{h}/N_{h}$ involve infectious hosts. 

Note that in the stationary points
$$
I_{h}^{*}=I_{v}^{*}=0\,,\quad
N_h^{*}=N_h=S_{h}^*+R_{h}^{*}
\quad\text{and}\quad
N_{v}^{*}=S_{v}^{*}=\widetilde{C}\,,
$$
where $\widetilde{C}=(1-(\mu/\alpha))\,C$ is the effective carrying capacity. 

\begin{figure}[!h]
\includegraphics[width=14cm,height=11cm]{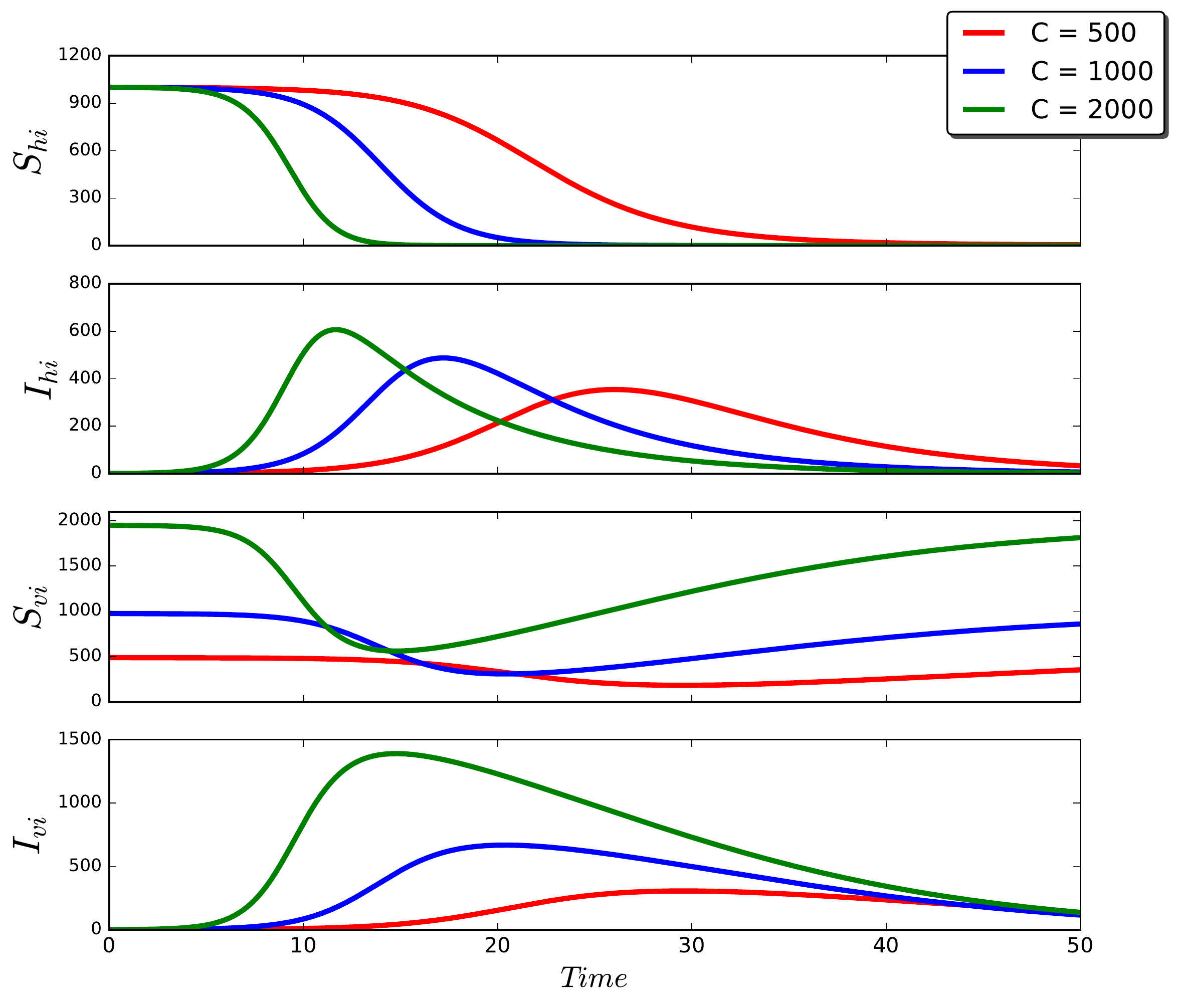}
\caption{{\bf Time series of the model \eqref{local model eq1}-\eqref{local model eq5} for different values of the carrying capacity $C$.}}
\label{fig:local model example}
\end{figure}

In order to illustrate the vector-borne disease dynamics for a single patch, we show in Figure \ref{fig:local model example} the time series of each one of the variables of the system \eqref{local model eq1}-\eqref{local model eq5} by selecting the parameter values given in Table \ref{table1}. In regards of the carrying capacity parameter $C$, we select the following values: $500$, $1000$ and $2000$. For each of these ones, we select as initial condition of the system the values $[S_{ho},I_{ho},R_{ho},S_{vo},I_{vo}] = [1000,1,0,\widetilde{C},0]$. 

\subsubsection*{Multiple connected patches model with hosts dwell time}

In this section we extend the local model described by equations \eqref{local model eq1}-\eqref{local model eq5} to the setting of multiple connected patches. Human mobility among distinct patches of $\Omega$ play an important role in the propagation of infectious diseases, and specifically of vector-borne ones. Consider an scenario where resident hosts of the $i$-th patch spend a period of their day (relevant with respect to vector activity) over the $j$-th patch. If there are infectious vectors in patch $j$, and some of them bite someone of these susceptible visitants, then these human hosts will bring the disease to patch $i$. In this way, some vectors in patch $i$ may also get infected, which at the same time will potentially infect, in subsequent days, both resident hosts in the $i$-th patch and visitors coming from other patches. On the other hand, infected visitants from $i$ could infect susceptible vectors in patch $j$ in case they were bitten, producing same result in visited patch $j$. 

Consequently, realistic models of vector-borne diseases spread demand a deep enough comprehension of this phenomena. In ``small'' regions, as cities are, human mobility among neighborhoods is usually repetitive and frequent. For instance, people go to work and return home almost every day, or go to the cinema or visit a friend, and then come back home in the same day. Furthermore, most of urban mobility within a city is by far caused by day-trippers. It is also very important (as far as applications are concerned) that this kind of motion follows patterns that can be measured, predicted or estimated. 

In order to incorporate these ideas to our epidemiological model, we follow a Lagrangian approach. We introduce a set of parameters $\{p_{ij}\}_{i,j=1}^{N}$ with the following meaning: for each pair $i,j$ with $1\leq i,j\leq N$, $p_{ij}$ is simultaneously (see the appendix): 
\begin{itemize}
\item The average fraction of people from patch $i$ that is in patch $j$ at any time. 
That means that the number of individuals from patch $i$ who are in patch $j$ is $p_{ij}N_{hi}$, at any time. 
\item The probability, at any time, to find a host from patch $i$ in patch $j$. 
\item The average fraction of time (a day or the corresponding time unit measure) that people from patch $i$ spend in patch $j$. 
That is why it is also called host dwell time  \cite{Bichara-CastilloChavez2016}. 
\end{itemize}
Furthermore, as $p_{ij}$ represent fractions we have the following conditions on mobility parameters: 
$$
0\leq p_{ij}\leq 1
\qquad\text{and}\qquad
\sum_{j=1}^{N}p_{ij}=1\,.
$$
In terms of this quantities, the number of individuals who are in patch $j$ is given by
\begin{equation}
w_j = \sum_{i=1}^{N} p_{ij}\,N_{hi}
\end{equation}
at any time $t$, regardless the patch where they are coming from (that is, including both own residents and day-trippers).  

Now we are ready to write the global model of multiple connected patches. Starting from the local model, namely \eqref{local model eq1}-\eqref{local model eq5}, we couple each local dynamic taking into account human mobility in the terms discussed before, obtaining the following system of $5N$ differential equations: for each $i=1,\ldots,N$
\begin{eqnarray}
\dot{S}_{hi} & = & 
-\sum^{N}_{j=1}\beta\,I_{vj}\,\frac{p_{ij}\,S_{hi}}{w_{j}}\,, 
\label{global model eq1} \\ 
\dot{I}_{hi} & = & 
\sum^{N}_{j=1}\beta\,I_{vj}\,\frac{p_{ij}\,S_{hi}}{w_{j}} 
-\gamma\,I_{hi}\,, 
\label{global model eq2} \\ 
\dot{R}_{hi} & = & 
\gamma\,I_{hi}\,, 
\label{global model eq3} \\ 
\dot{S}_{vi} & = & 
\alpha_{i}\,N_{vi}\,\left(1-\frac{N_{vi}}{C_{i}}\right) 
-\sum^{N}_{j=1}\beta\,S_{vi}\,\frac{p_{ji}\,I_{hj}}{w_{i}} 
-\mu_{i}\,S_{vi}\,, 
\label{global model eq4} \\ 
\dot{I}_{vi} & = & 
\sum^{N}_{j=1}\beta\,S_{vi}\,\frac{p_{ji}I_{hj}}{w_{i}} 
-\mu_{i}\,I_{vi}\,. 
\label{global model eq5} 
\end{eqnarray}

With respect of equations \eqref{global model eq1} and \eqref{global model eq2}. For a fixed $i$, for each $j$ with $1\leq j\leq N$, $\beta\,I_{vj}$ is the total number of bites of infectious vectors to hosts per day in $j$. From which, only a portion are on susceptible ones to become sick in the $i$-th patch. Since $p_{ij}$ is the average fraction of people from patch $i$ in patch $j$, and assuming this average is independent of the susceptible or infectious populations, then $p_{ij}\,S_{hi}$ is the number of susceptible hosts from patch $i$ who are in patch $j$. Thus, if $w_{j}$ is the number of people in patch $j$, the portion of susceptible hosts from patch $i$ among people in patch $j$ is $p_{ij}\,S_{hi}/w_{j}$. According to these interpretations, $\beta\,I_{vj}\,p_{ij}\,S_{hi}/w_{j}$ is the rate of new infected hosts from patch $i$ being infected in patch $j$. Finally, we add over all patches $j$ to get the total number of infected hosts from patch $i$. 
 
Analogously, in equations \eqref{global model eq4} and \eqref{global model eq5}. $\beta\,S_{vi}$ is the number of bites performed by susceptible vectors in patch $i$, $p_{ji}\,I_{hj}$ is the number of infective human visitors of patch $i$ coming from patch $j$, and $p_{ji}\,I_{hj}/w_{i}$ is the fraction of infectious hosts from patch $j$ in patch $i$. Consequently, $\beta\,S_{vi}\,p_{ji}\,I_{hj}/w_{i}$ is the number of infected vectors per unit time in patch $i$ caused by human hosts from patch $j$. Finally, we add over all patches $j$ to get the total number of infected vectors in patch $i$. 

The entire mobility configuration is determined by the coupling matrix $P=(p_{ij})_{i,j=1}^{N}$, which is a right stochastic matrix (its entries are probabilities and its rows sum one) non necessarily symmetric (\textit{i.e.} in general $p_{ij}\neq p_{ji}$).

By way of example, we perform a numerical simulation for a network of $N =5$ patches coupled in the following configuration: 
\begin{equation}\label{eq:P-example1}
P_{e.g.1} = \begin{pmatrix}
0.4143 & 0.2922 & 0.2934 & 0 & 0\\
0.3340 & 0.3597 & 0      & 0.3061 & 0\\
0.3773 & 0      & 0.6226 & 0 & 0\\
0      & 0.6658 & 0      & 0.1307 & 0.2034\\
0      & 0      & 0      & 0.3954 & 0.6045
\end{pmatrix}\,.
\end{equation}

For these examples we choose the same parameter values given in Table \ref{table1}, but we vary the carrying capacity for each patch taken from a power law probability distribution in the range $C_{i} \in [100,1000]$. In regards of the initial conditions, we assume that at the initial time there are not infectious host and vectors; \textit{i.e.} $I_{hi}(t=0) = I_{vi}(t=0) = 0$ for almost all values of the index $i$. In this example we select randomly a single patch and we introduce an infected host, that is $I_{hj}(t=0)=1$ for a given randomly selected patch $j$.
The initial number of susceptible hosts $S_{hi}(t=0) \in [500,1000]$ where taken from a power law distribution and $S_{vi}(t=0)=C_{i}(1-\mu_{i}/\alpha_{i})$ for all $i$. The rest of the variables start at zero.


\begin{figure}[!h]
\includegraphics[width=14cm,height=11cm]{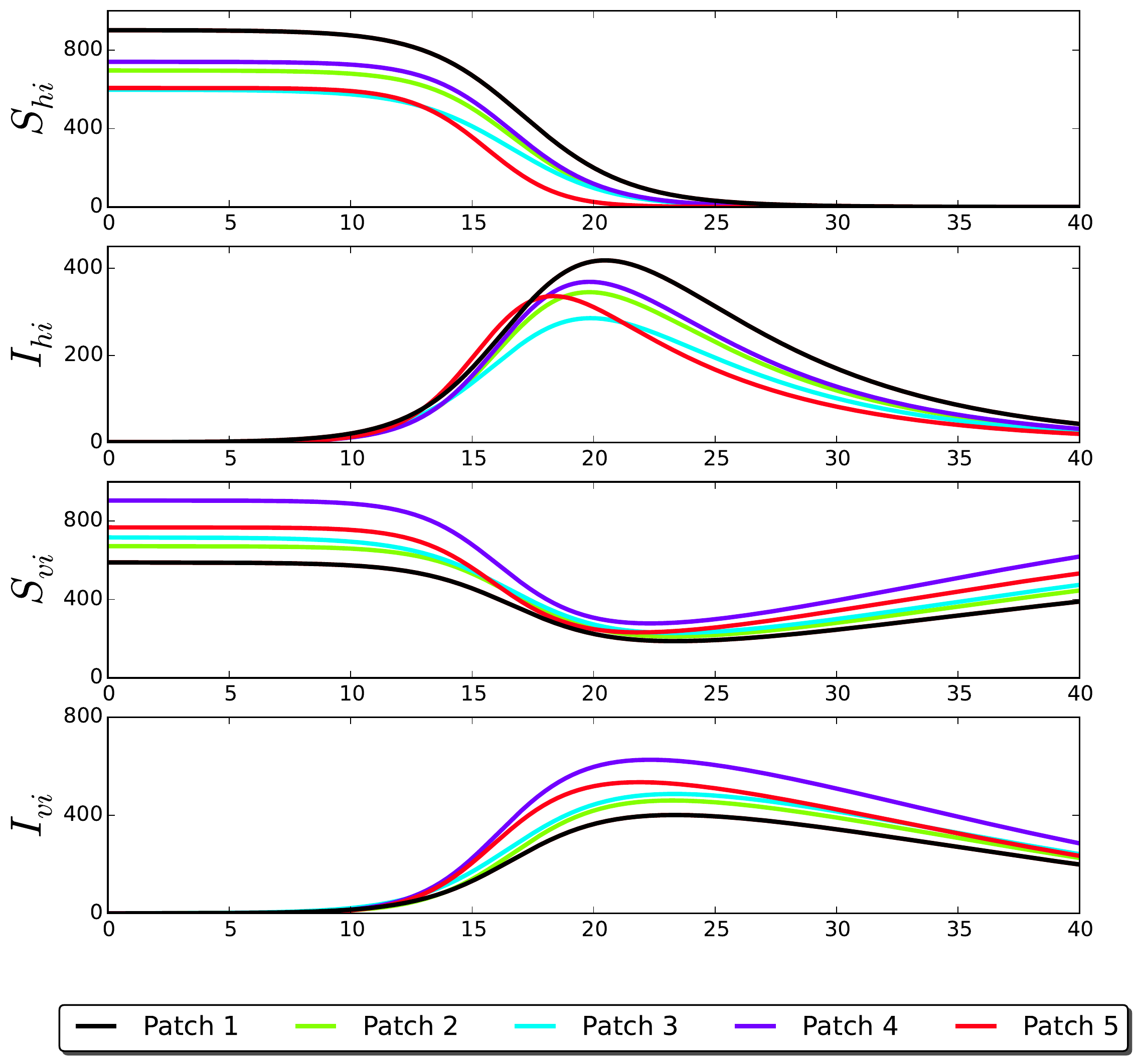}
\caption{{\bf Time series of the multiple patch connected according to the dwell-time matrix $P_{e.g.1}$ given in \eqref{eq:P-example1}.}}
\label{fig:global model example}
\end{figure}


\section*{Results}

\subsection*{Expressions for the indexes}

In order to find an explicit expression for the secondary human cases $R_{kj}^{(v)}$ caused by a single infected mosquito in $k$, we have to multiply the rate of infections generated from mosquitoes of patch $k$ to visitors from patch $j$, i.e. $\frac{\beta\,S_{hj}\,p_{jk}\,I_{vk}}{w_{k}}$ times the characteristic duration time that a mosquito remains infected, $1/\mu_{k}$. Then we evaluate this quantity with a single infected vector in the disease-free equilibrium. That is
\begin{equation}
R_{kj}^{(v)} = 
\left.\frac{\beta\,S_{hj}\,p_{jk}\,I_{vk}}{\mu_{k}\,w_{k}}\right|_{S_{hj}=N_{hj},I_{vk}=1}
\,=\frac{\beta\,N_{hj}\,p_{jk}}{\mu_{k}\,w_{k}}\,.
\end{equation}

In a similar way, from the proposed model an explicit expression for secondary vector infections in patch $k$ that were caused by travelers from patch $i$ at the beginning of an epidemic is
\begin{equation}
R_{ik}^{(h)} = 
\frac{\beta\,p_{ik}\,N_{vk}}{\gamma w_{k}}\,.
\end{equation}

Thus, $R_{i}^{(h)}$ is given by
\begin{equation}
R_{i}^{(h)} = 
\sum_{j=1}^{N} \sum_{k=1}^{N} R_{ik}^{(h)}\,R_{kj}^{(v)} = 
\frac{\beta^{2}}{\gamma}\,\sum_{k=1}^{N} \frac{p_{ik}\,N_{vk}}{\mu_{k}\,w_{k}}\,,
\end{equation}
and the transmission risk index becomes
\begin{equation}
TR_{i} = 
\frac{N_{hi}}{N_{h}}\,R_{i}^{(h)} = 
\frac{\beta^{2}\,N_{hi}}{\gamma N_{h}}\,\sum_{k=1}^{N} \frac{p_{ik}\,N_{vk}}{\mu_{k} w_{k}}\,.
\end{equation}

In a similar way the vulnerability index takes the form
\begin{equation}
VR_{j} = 
\sum_{i=1}^{N}\sum_{k=1}^{N} \frac{N_{hi}}{N_{h}}\,R_{ik}^{(h)}\,R_{kj}^{(v)} = 
\frac{\beta^{2} N_{hj}}{\gamma N_{h}}\,\sum_{k=1}^{N} \frac{p_{jk} N_{vk}}{\mu_{k} w_{k}}\,,
\end{equation}
and finally the vector transmission risk is
\begin{equation}
VTR_{i} \simeq 
\frac{w_{i}}{N_{h}}\,\left(1 - e^{-N_{vi}/w_{i}} \right)\,R_{i}^{(v)}\,,
\end{equation}
where
\begin{equation}
R_{i}^{(v)} = \sum_{j=1}^{N} R_{ij}^{(v)} = 
\frac{\beta}{\mu_{i}}\,.
\end{equation}

\subsection*{Numerical Simulations}

In order to show how the risk indexes can be used to guide control strategies we simulate an ensemble of Dengue outbreaks using the model \eqref{global model eq1}-\eqref{global model eq5}. First we perform an ensemble of 200 simulations with parameter values shown in Table \ref{table1}. For this example we used $5$ patches, the number of humans and the carrying capacity in each patch was taken randomly. The data of each patch are shown in the Table \ref{tableD}, where $N_{hi}$ is the number of humans in patch $i$, $N_{vi}$ is the number of vectors in patch i. 

\begin{table}[!ht]
\begin{adjustwidth}{-2.25in}{0in} 
\caption{
{\bf Parameter values for each path.}}
\begin{tabular}{crrr}
\hline
   Patch &   $C_i$ &   $N_{hi}$ &   $N_{vi}$  \\
\hline
       1 & 1435.90  &       6587 &       1400  \\
       2 & 2564.10  &       7292 &       2500  \\
       3 & 1025.64 &       9791 &       1000 \\
       4 & 2051.28 &       5875 &       2000  \\
       5 & 1025.64 &       4354 &       1000  \\
\hline
\end{tabular}
\label{tableD}
\end{adjustwidth}
\end{table}

We used two types of mobility matrix, the first is an unrestricted mobility, the average fraction of people from patch $i$ can go to any other patch $j$ and the  specific value for the plotted examples is

\begin{equation} P_1 = \left(\begin{matrix}
 0.2967 & 0.0765&  0.1650 & 0.0642&  0.3973\\
  0.1840 &  0.1304&  0.2441&  0.2919&  0.1495\\
  0.2283&  0.1299&  0.1900&  0.2244&  0.2271\\
  0.0843&  0.1977 &  0.2613&  0.2146&  0.2418\\
  0.2609&  0.2294&  0.1724&   0.2063 & 0.1307\\
\end{matrix}\right)\nonumber
\end{equation}

The second type of tested mobility matrix resembles a network of Barabasi-Albert type. In this kind of mobility, people from a particular patch travel only to some other patches but not all. The particular mobility matrix used for the plotted examples is

\begin{equation} P_2 = \left(\begin{matrix}
  0.2127 & 0.2633&  0         & 0.5238 & 0        \\
  0.2675 & 0.2175&  0.1698 & 0         & 0.3450\\
  0         & 0.2074&  0.7925&  0        & 0        \\
  0.7656 & 0        &  0        &  0.2343 & 0       \\
  0        & 0.3662&  0       &  0       & 0.6337\\
\end{matrix}\right)\nonumber\end{equation}

\begin{figure}[!h]
\centering
\subfigure[
]
{\includegraphics[width=0.5\textwidth]{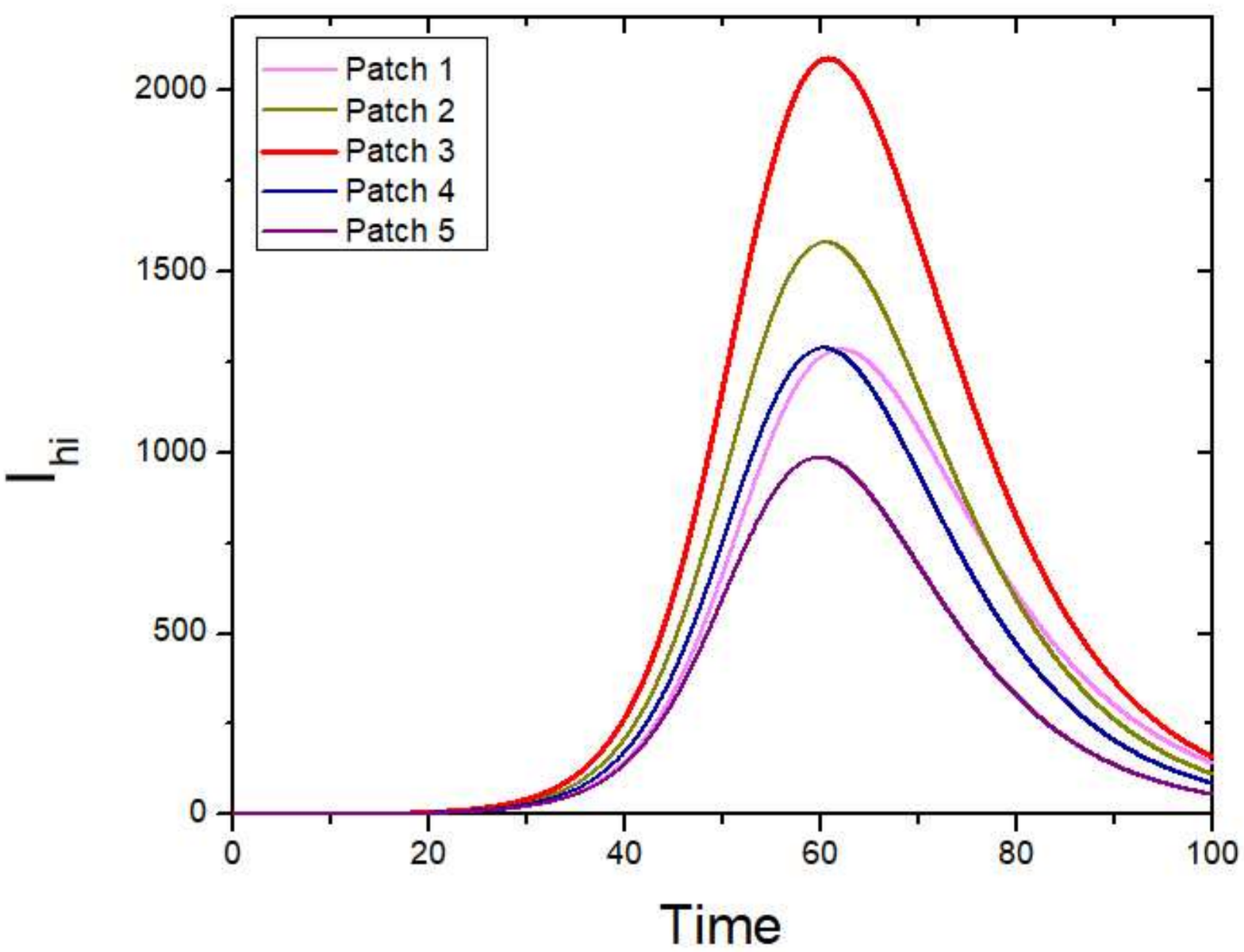}}%
\subfigure[
]
 {\includegraphics[width=0.5\textwidth]{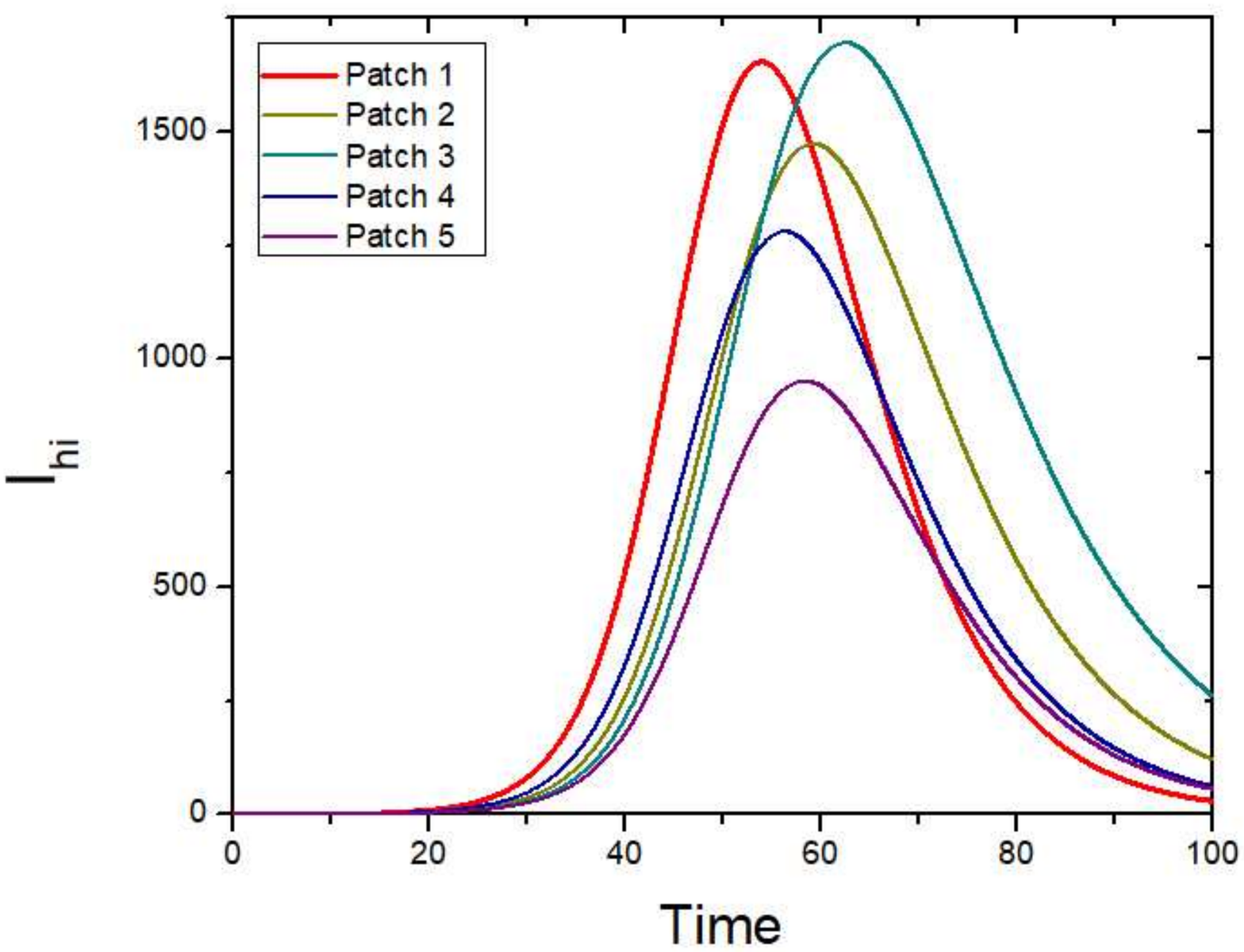}}
\caption{Dynamics of the number of infected humans per patch in a network of 5 nodes for a fully connected network (a) and a Barabasi-Albert network (b). In (a) the vulnerability index $VR_{i}$ indicate patch 3 (red line) as the more vulnerable and in (b) the more vulnerable as indicated by $VR_{i}$ is patch 1 (red line). The more vulnerable patch corresponds to the one with more cases at the beginning of the epidemic.}  
\label{ejemploW}
\end{figure}

The patch with greater vulnerability index $VR_{i}$ corresponds to the patch with more infections at the beginning of the outbreak as can be seen in Fig. \ref{ejemploW}.

\begin{figure}[!h]
\centering
\subfigure[
]
{\includegraphics[width=0.5\textwidth]{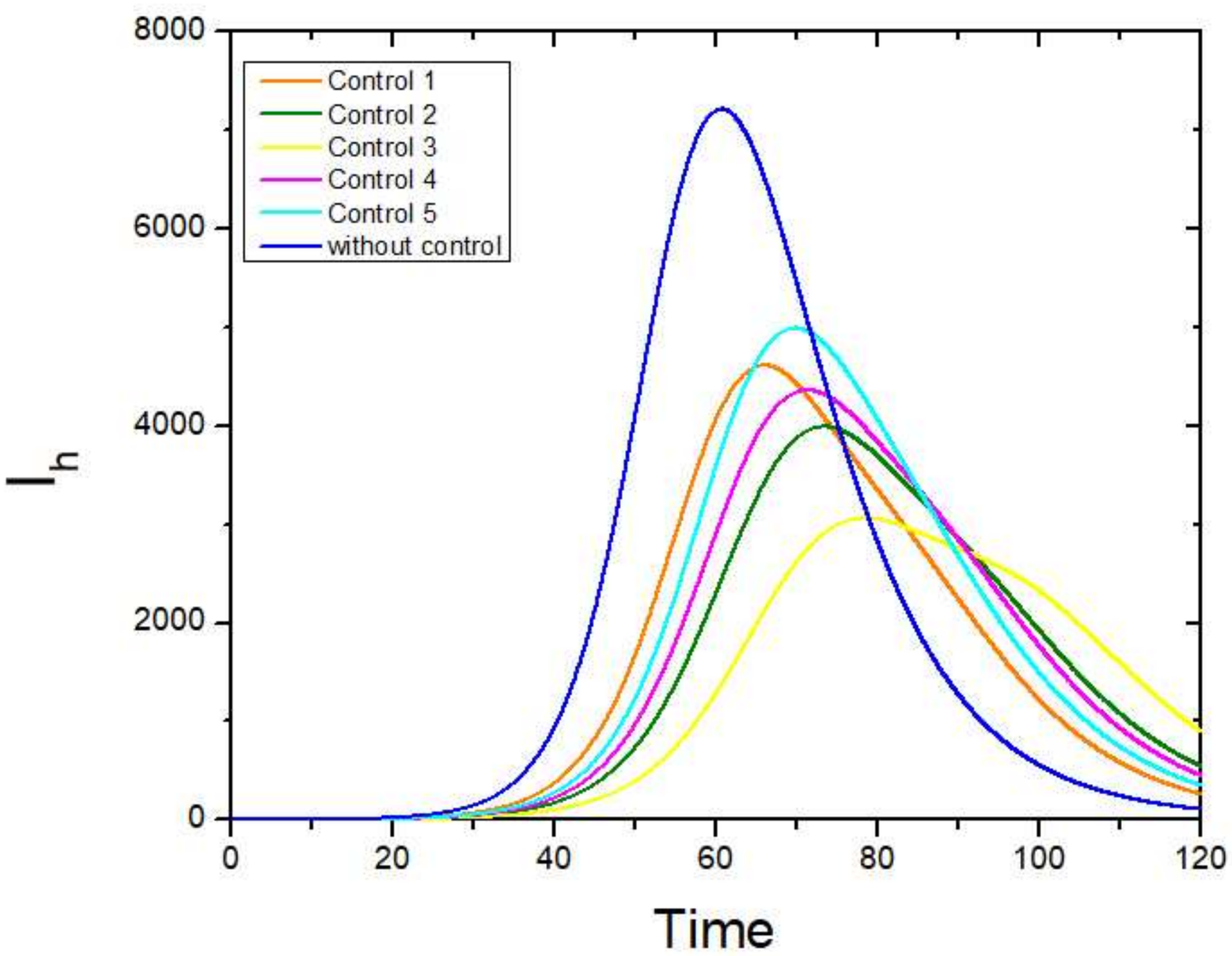}}%
\subfigure[
]
{\includegraphics[width=0.5\textwidth]{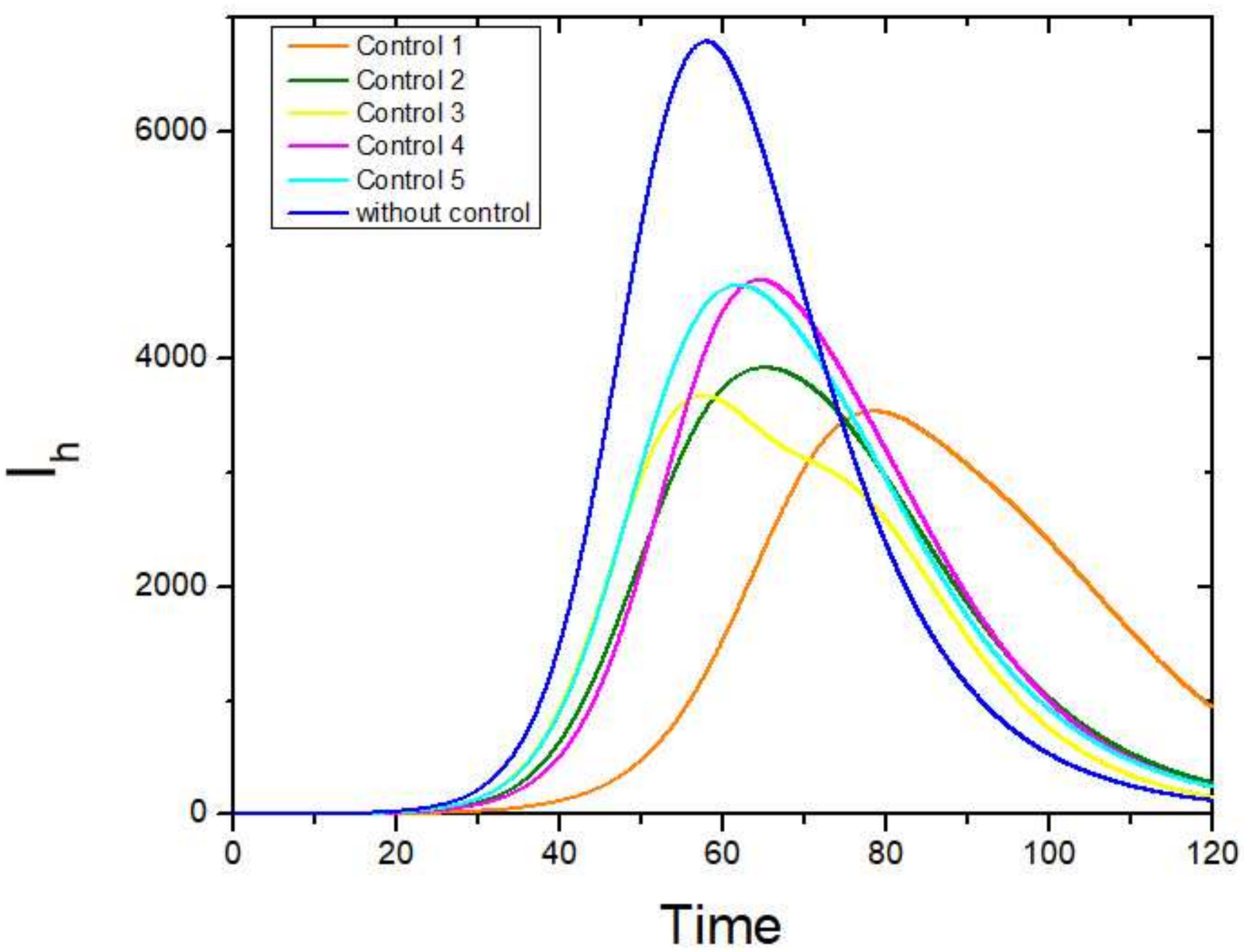}}
\caption{Dynamics of the infected population system-wide with control applied in different patches. (a) In a fully connected network the transmission index $TR_{i}$ is greater in patch 3. (b) In a Barabasi-Albert network the patch with more transmissibility is number 1 as indicated by its $TR_{i}$. The more effective strategy is to reduce the infected hosts (by treatment and hospitalization) in the patch with greater $TR_{i}$.} 
 \label{ejemplo2R}
\end{figure}

We then compare the effect of apply an specific control measure in a randomly selected patch with the effect of applying the same control measure but in the patch indicated by an adequate index. The first tested control measure was the fast isolation of infected people by hospitalizing them as soon as an infection is detected. We simulate this by increasing the recovery rate $\gamma$ of a single patch. Fig. \ref{ejemplo2R} shows as expected, that on average this measure is more effective when applied in the patch with greater transmission index $TR_{i}$ than in any other patch. 

\begin{figure}[!h]
\centering
\subfigure[
]
{\includegraphics[width=0.5\textwidth]{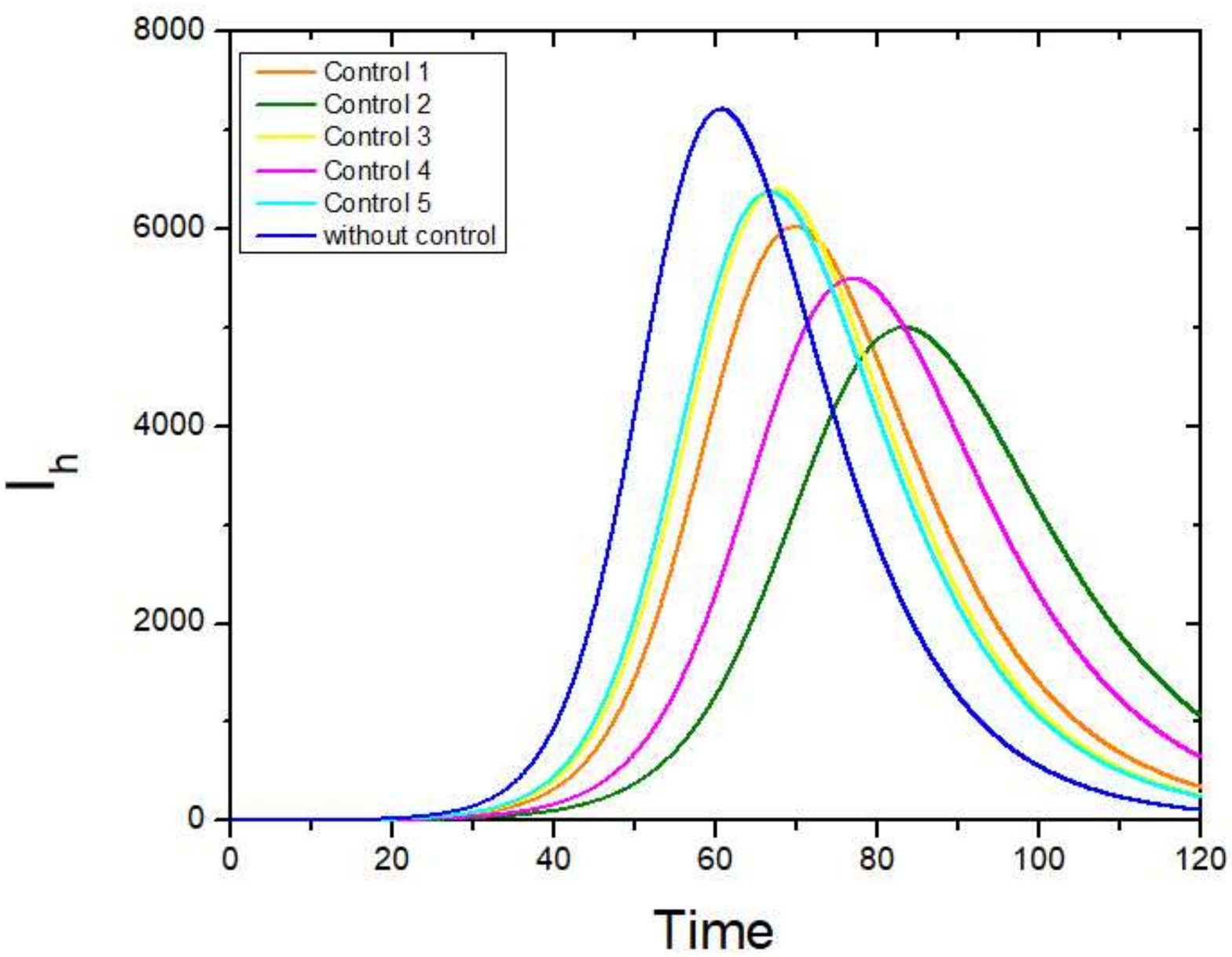}}%
\subfigure[
]
{\includegraphics[width=0.5\textwidth]{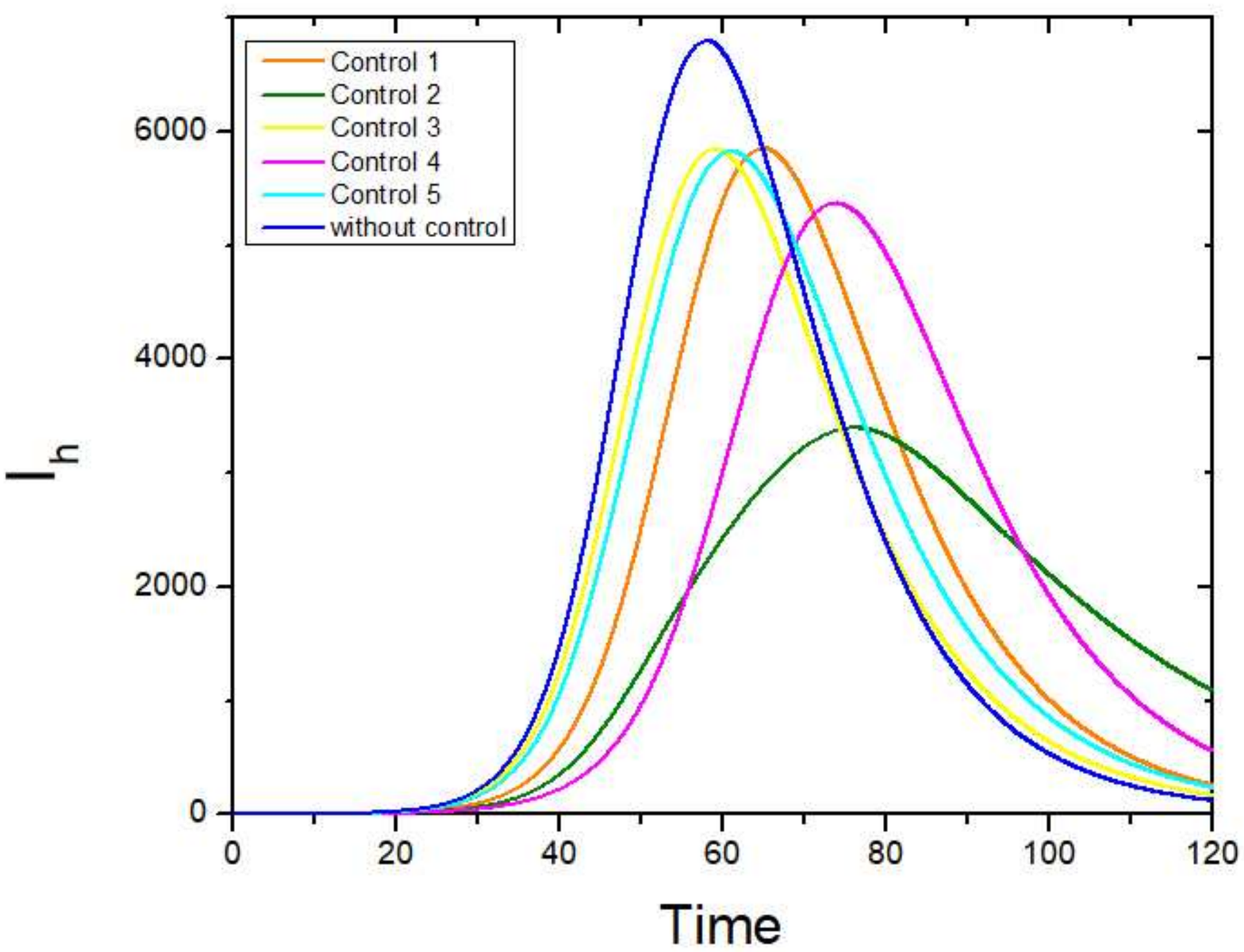}}
\caption{Dynamics of the infected population system-wide with control applied in different patches. (a) In a fully connected network the vector transmission index $VTR_{i}$ is greater in patch 2. (b) In a Barabasi-Albert network the patch with more vector transmissibility is number 4 as indicated by its $VTR_{i}$. The more effective strategy at the beginning of the outbreak is to reduce the carrying capacity in the patch with greater $VTR_{i}$.}
\label{ejemplo2C}
\end{figure}

Finally the vector transmission index $VTR_{i}$ was able to indicate the patch where the reduction of the local carrying capacity led to a more effective strategy to reduce the initial growth of the epidemic (Fig. \ref{ejemplo2C}).

\section*{Discussion}

We propose three risk indexes for the transmission of vector-borne diseases. Two of them are related with the transmission risk of patches and the other one gives a measure of the vulnerability of each patch. It is worth to notice their skill to localize epidemiological risk at different nodes of a net. 

The transmission indexes quantify the potential secondary cases that an individual or a vector of a given patch can produce system-wide. On the other hand, the vulnerability index quantify the potential secondary cases that an infected individual in the net can cause in the target patch.

Specifically, the transmission risk index $TR_{i}$ is the probability of a host in patch $i$ became the first infected host, times secondary human infections they would potentially cause. The second index $VTR_{i}$, called vector transmission risk index, is defined as the probability of a vector gets first infected in patch $i$, times secondary human infections it would potentially cause. The vulnerability index $VR_{i}$ is secondary host infections in patch $i$ averaged over a randomly distributed initial infected human in the system. 

These risk indexes can be used in different ways. First, they indicate optimal places to apply prevention measures. They are also useful to guide the monitoring campaigns for an early detection of epidemics. And finally, they can guide control and mitigating strategies at the beginning of an outbreak. 

For example, the transmission risk index $TR_{i}$ can guide in a repellent distribution campaign in order to reduce the contagion probability of the greatest dispersers. Another control strategy is to isolate by hospitalization infected individuals of such a neighborhood. The vector transmission risk $VTR_{i}$ indicates the most suitable places where abatization and fumigation campaigns will be more effective. On the other hand, the vulnerability risk index $VR_{i}$ can guide investments in health clinics. That will increase the public health capacity in this particular area where many cases will likely arise. In addition to this, as it is expected to find more cases in the neighborhood with highest $VR_{i}$, this is a good place for incidence monitoring. That will help in detecting increased anomalous endemic states or an outbreak. 

In order to illustrate these control measures, we compared a set of simulations in the frame of a proposed particular model for vector-borne diseases spread in a multi-patch system for a Dengue epidemics. In addition to vector and host densities, this model takes into account human mobility. We have observed that an immediate medical attention gives benefits to the whole population if the neighborhood with highest transmission risk $TR_{i}$ is prioritized. In contrast, reducing the hatcheries for example by abatization brings more benefits if the neighborhood with highest vector transmission risk $VTR_{i}$ is prioritized. 


\section*{Acknowledgments}

The authors thankfully acknowledge the computer resources, technical expertise and support provided by the Laboratorio Nacional de Supercómputo del Sureste de México, CONACYT network of national laboratories.
\nolinenumbers

%

\bibliographystyle{plos2015.bst}
\bibliography{Bibliography}

\section*{Appendix: concerning host dwell times}
\label{appendixB}
Let us clarify the meaning of parameters $\{p_{ij}\}_{i,j=1}^{N}$. Fix an index $i$ with $1\leq i\leq N$, and consider a typical human individual of patch $i$. Assign them provisionally a second index $k$, with $1\leq k\leq N_{hi}$. Their motion can be tracked down with a function $\phi_{i,k}\,:\,[0,\infty)\longrightarrow\{1,\ldots,N\}$ in the following way: for any $t\geq 0$ and $j$ ($1\leq j\leq N$) $\phi_{i,k}(t)=j$ means that individual $k$ from patch $i$ is in patch $j$ at time $t$. Denote $\Phi_{i,j}(t)$ the number of human individuals of patch $i$ who are in patch $j$ at time $t$ (number of indexes $k$ such that $\phi_{i,k}(t)=j$). 

If the population is large enough and their mobility habits are repetitive, it makes sense to assume that, in spite of functions $\phi_{i,k}$ depend on time, quantities $\Phi_{i,j}$ actually do not. Denote $p_{ij}=\Phi_{i,j}/N_{hi}$ for any pair $i,j$. Equipped with this convention, clearly $p_{ij}$ is the average fraction of people from patch $i$ in patch $j$ at any time, and $\Phi_{i,j}=p_{ij}N_{hi}$ is the number of individuals from patch $i$ who are in patch $j$, at any time. On the other hand, given an individual from patch $i$ (given $k$ with $1\leq k\leq N_{hi}$), the probability to find this host in some particular patch $j$ is the probability of having $\phi_{i,k}(t)=j$. Assuming this one to be independent of the particular $k$ (population is homogeneously mixed at every single patch), this quantity is obviously number of favorable cases over total cases, \textit{i.e.} $\Phi_{i,j}/N_{hi}=p_{ij}$. 

The third interpretation, in terms of dwell times, is a little more involved. Let's denote $\chi_T$ the indicator function of the set $T$ (\textit{i.e.} $\chi_T(t)=1$ if $t\in T$ and $\chi_T(t)=0$ elsewhere), and $|T|$ the length (or Lebesgue linear measure) of $T$. Suppose functions $\phi_{i,k}$ are simultaneously $\tau$-periodic (as mobility habits are repetitive) for some $\tau>0$ (the time unit measure). Thus $\phi_{i,k}=\sum_{j=1}^{N}j\,\chi_{T_{ijk}}$ (restricted to its period) for some sets $\{T_{ijk}\}$ in $[0,\tau]$. This is equivalent to say that $\phi_{i,k}(t)=j$ iff $t\in T_{ijk}$, so $|T_{ijk}|/\tau$ is the fraction of time that individual $k$ from patch $i$ spends in patch $j$ in any period. Also, we find that
$$
\Phi_{i,j}(t)=
\sum_{
\begin{subarray}{c} 
k=1 \\ 
\phi_{i,k}(t)=j 
\end{subarray}
}^{N_{hi}}1=
\sum_{k=1}^{N_{hi}}\chi_{T_{ijk}}(t)\,.
$$
Thus, finally, the average fraction of time that people from patch $i$ spend in patch $j$ is
$$
\frac{1}{N_{hi}}\,\sum_{k=1}^{N_{hi}}\frac{|T_{ijk}|}{\tau}=
\frac{1}{N_{hi}\,\tau}\,\int_{0}^{\tau}\Phi_{i,j}(t)\,dt=
\frac{p_{ij}\,N_{hi}\,\tau}{N_{hi}\,\tau}=
p_{ij}\,.
$$
Furthermore, we have the following conditions on mobility parameters. Clearly $0\leq\Phi_{ij}\leq N_{hi}$ and $\sum_{j=1}^{N}\Phi_{i,j}(t)=N_{hi}$, so we have
$$
0\leq p_{ij}\leq 1
\qquad\text{and}\qquad
\sum_{j=1}^{N}p_{ij}=1\,,
$$
and we also find that the number of individuals who are in patch $j$ is
$$
w_j=
\sum_{i=1}^{N}\Phi_{i,j}=
\sum_{i=1}^{N}p_{ij}\,N_{hi}
$$
at any time $t$, regardless the patch where they are coming from (that is, including both own residents and day-trippers).Of course, this way of reasoning is merely motivational. We admit any real value between zero and one for $p_{ij}$, and not only $p_{ij}=\nu/N_{hi}$, with $\nu,N_{hi}$ integers fulfilling $0\leq\nu\leq N_{hi}$.

\end{document}